\documentclass[11pt]{article}
\usepackage[dvipdfmx]{graphicx}
\usepackage{color}
\begin{document}

\begin{center}{\Large Extended Relative Power Contribution \\[2mm]that Allows to Evaluate the Effect of Correlated Noise}\\

\vspace{8mm}
{\large Genshiro Kitagawa}\\[1mm]
The Institute of Statistical Mathematics\\[-1mm]
and\\[-1mm]
Graduate University for Advanced Study\\[2mm]
{\large Yoko Tanokura}\\[1mm]
CIPPS Inc., Visiting Fellow\\[2mm]

{\large Seisho Sato}\\[1mm]
Graduate School of Economics, The University of Tokyo

\vspace{5mm}
{\today}

\end{center}

\noindent{\bf Abstract}

We proposed an extension of Akaike's relative power contribution that could be applied to data with correlations between noises. This method decomposes the power spectrum into a contribution of the terms caused by correlation between two noises, in addition to the contributions of the independent noises. Numerical examples confirm that some of the correlated noise has the effect of reducing the power spectrum.

\vspace{1mm}
\noindent

\section{Introduction}
Multivariate dynamic systems, such as industrial plants like cement rotary kilns and thermal power plant boilers, ships navigating in the ocean, and economic systems, vary with many variables with serial and time series correlations. Akaike (1968) proposed the concept of relative power contribution as a tool to analyze multivariate systems with such feedback. This method has been applied to the analysis and control of many real systems and many successful examples have been reported, for example, analysis and statistical control of a cement rotaly kilns and boiler of electric power plant, design of autopilot sytem of ships, and analysis of financial asset prices. 
For more details, readeres are referred to Akaike and Nakagawa (1988) and Akaike and Kitagawa (1998).

The relative power contribution is defined under fairly strong assumption that the variance-covariance matrix of the noise in a multivariate AR model representing a multivariate time series system is diagonal. Therefore, when there is a strong correlation between the noise terms in each variable, as in economic systems, the results obtained by assuming uncorrelation among the noises may have a large bias, making the application of this concept problematic.

Therefore, extending Akaike's concept of relative power contribution to include the case where there is correlation between noises is an important issue, as it allows for application to diverse fields of multivariate systems.
On this issue, Tanokura and Kitagawa (1994) defined a generalized relative power contribution that can be applied when the uncorrelatedness of noise does not hold by decomposing the $k\times k$ variance-covariance matrix of noise into the sum of $k(k+1)/2$ rank 1 matrices. The distinctive feature of this method is that it generally uses not only the correlation of two noise variables, but also the correlated noise of three or more variables. This method is mainly used in the analysis of economic systems.
See Tanokura and Kitagawa (2015) for details.

In this paper we consider a simpler method than the above method to extend Akaike's power contribution. When the variance-covariance matrix of the noise contains off-diagonal elements, the power spectrum of each component requires a term related to the correlation coefficient between the two noises, in addition to the contribution from the independent noises. By considering this as a contribution to the power spectrum from the correlated noise, we can extend the concept of relative power contribution. In this case, however, some of the contribution terms take negative values. The meaning of this negative contribution will be discussed through examples of actual time series and simulations.

This paper consists of seven sections: in Section 2, we review Akaike's concept of the relative power contribution and briefly introduce the generalized relative power contribution by Tanokura and Kitagawa (1994).
Section 3 defines the new extended relative power contribution.
Two examples are presented to illustrate this approach.
In Section 4, we consider the generalized relative power contribution for ship's motions. In the case of this system, the correlations between the noises are relatively small, and in this case the results are similar to the Akaike relative contribution, so the application of the extended relative power contribution is not essential.
Section 5, on the other hand, analyzes GDP data, which yields quite different results since there are strong correlations between noise inputs.
Chapter 6 reinterprets the results of the extended relative power contribution from a simulation perspective. Chapter 7 discusses the concluding remarks.

\section{A Review of Akaike's Relative Noise Contribution}
Assume that $y_n$ is $k$-dimensional time series and consider a multivariate autoregressive model
\begin{eqnarray}
y_n =\sum_{j=1}^m A_j y_{n-j} + v_n,
\end{eqnarray}
where $A_j$ is $k\times k$ autoregressive coefficient matrix for lag $j$, 
$v_n$ is $k$-dimensional Gaussian white noise with mean 0 and variance-covariance 
matrix $V$, $N(0,V)$.
In the definition of Akaike's relative power contribution (or relative noise contribution),
it is assumed that the variance covariance matrix of the noise input $V$ is diagonal and
is given by $V = \mbox{diag}(\sigma^2_1,\ldots ,\sigma^2_k)$.

Then the cross spectrum matrix of the AR process is given by 
\begin{eqnarray}
  P(f) = A(f)^{-1}V(A(f)^{-1})^{\ast} \equiv B(f)VB(f)^{\ast}, \quad -1/2 \leq f \leq 1/2,
  \label{Eq-Cross-spectrum}
\end{eqnarray}
where $\ast$ denote the complex conjugate of the compex matrix and $A(f)$ is the Fourier
transform of the AR coefficient matrices defined by
\begin{eqnarray}
  A(f) = I - \sum_{j=1}^m A_j e^{-2\pi ijf}, \quad -1/2 \leq f \leq 1/2.
\end{eqnarray}
Therefore, from equation (\ref{Eq-Cross-spectrum}), the power spectrum of $j$-th time series is given by
\begin{eqnarray}
  p_{jj} = \sum_{l=1}^k b_{jl}(f)\sigma^2_l b_{jl}^{\ast}(f) = \sum_{l=1}^k |b_{jl}(f)|^2\sigma^2_l ,
\end{eqnarray}
where $b_{jl}$ is the ($j,l$) element of the complex matrix $B(f)$.
This indicates that the power spectrum can be decomposed into $k$ componoents where
$|b_{jl}(f)|^2\sigma^2_l$ can be considered as the contribution from the noise input
to the $l$-th time series.
Based on this consideration, Akaike (1969) defined the \textbf{relative power contribution} (relative noise contribution) to $j$ from $l$ at frequency $f$  by
\begin{eqnarray}
  r_{jl} = \frac{|b_{jl}(f)|^2\sigma^2_l}{p_{jj}(f)}, \quad l=1,\ldots ,k.
\end{eqnarray}

Relative power contribution is useful in the analysis of feedback systems and the design of optimal control systems, and have been applied in many areas, including cement rotary kilns (Akaike and Nakagawa (1989)), thermal power plants (Nakamura and Akaike (1981), Akaike and Kitagawa (1998)), ships motion and engine system (Ohtsu (1912) and Ohtsu, Peng and Kitagawa (2015)), and macro economics (Tanokura and Kitagawa (2015)). 

In this paper, we will refer to 
\begin{eqnarray}
  s_{jl} = |b_{jl}(f)|^2\sigma^2_l,  \quad l=1,\ldots ,k
\end{eqnarray}
as the \textbf{absolute power contribution} of $l$-th time series to the $j$-th time series at frequency $f$.
As will be seen in the examples in the next section, important information can be obtained from the absolute power contribution.

Tanokura and Kitagawa (1994, 2015) extend Akaike's power contribution by
decomposing the variance-covariance matrix in the following manner:
\begin{eqnarray}
  V = \sum_{j=1}^k \alpha_{jj}I_{H_j(k-1)}I^T_{H_j(k-1)}
    + \sum_{m=0}^{k-2}\sum_{j=1}^{m+1} \alpha_{k-(m+1)+j,j}I_{H_j(k)}I^T_{H_j(k)},
\end{eqnarray}
where $I_{H_j(m)} = \{i_{jm}(1),\ldots ,i_{jm}(k)\}$ ia an $k$-dimensional vector.
of which $m$ components are 0 and $(k-m)$ components are either 1 or $-1$,
depending on the signs of correlations for $m=0,\ldots ,k-1;j=1,\ldots ,m+1$.
Here $H_j(k)$, the sufix of $I_{H_j(m)}$, is a subset of $H_j(k)=\{h_{j,1},\ldots ,h_{j,m}\}$ of $H=\{1.\ldots ,k\}$ and indicates the components of 0 of $I_{H_j(m)}$.

Then the power spectrum of the $r$-th component is expressed as
\begin{eqnarray}
p_{rr}(f) = \sum_{j=1}^k \alpha_{jj}|b_{rj}(f)|^2
   + \sum_{m=0}^{k-2}\sum_{j=1}^{m+1} \alpha_{k-(m+1)+j,j}
     \sum_{h=1,h\neq r}^k \sum_{n=1,n\neq r}^k C_{rjm}(h)c_{rjm}(n)^{\ast},
\end{eqnarray}
where $c_{rjm}(h) = i_{jm}b_{rh}(f)$.
Finally, the generalized power contribution is defined as
\begin{small}\begin{eqnarray}
r_{rjm}(f) = \left\{ \begin{array}{ll}
     \displaystyle \frac{\alpha_{k-(m+1)+j,j}\sum_{h=1,h\neq r}^k \sum_{n=1,n\neq r}^k C_{rjm}(h)c_{rjm}(n)^{\ast}}{p_{rr}(f)},  &   m=0,\ldots ,k-2; j=1,\ldots ,m+1\\
     \displaystyle \frac{\alpha_{jj}|b_{rj}(f)|^2}{p_{rr}(f)},  & m=k-1; j=1,\ldots ,m
             \end{array}\right.
\end{eqnarray}\end{small}
The applications of this generalized power contribution can be found in
Tanokura and Kitagawa (2015).

\section{Noise Contribution in the Presence of Correlated Noise}
Although the relative power contribution is very useful, the assumption that the variance-covariance matrix of the noise is a diagonal matrix limits its applicability.
Here, we consider a way to extend the relative power contribution to the case where the AR model has a variance-covariance matrix in general form. As shown in the previous section, based on the same motivation, Tanokura and Kitagawa (2004) proposed a method by decomposing the variance-covariance matrix to $k(k+1)/2$ rank one matrices, but here we consider a method that can directly use the off-diagonal terms of the covariance matrix as they are.

Assume that the variance-covariance matrix of the noise term of the $k$-variate autoregressive model is given by
\begin{eqnarray}
  V = \left[ \begin{array}{ccc} 
           \tau_{11} & \cdots & \tau_{1k} \\
           \vdots    & \ddots & \vdots    \\
           \tau_{k1} & \cdots & \tau_{kk}
      \end{array}\right].
\end{eqnarray}
Then from equation (\ref{Eq-Cross-spectrum}), the power spectrum of the $j$-th time series, i.e., the $(j,j)$ component of the cross spectral matrix is obtained by
\begin{eqnarray}
   p_{jj}(f) &=& [b_{j1} \cdots b_{jk}] \left[\begin{array}{ccc}
           \tau_{11} & \cdots & \tau_{1k} \\
           \vdots    & \ddots & \vdots    \\
           \tau_{k1} & \cdots & \tau_{kk}
      \end{array}\right] 
           \left[ \begin{array}{c} b_{j1}^{\ast} \\ \vdots \\ b_{jk}^{\ast} \end{array} \right] \nonumber \\
    &=& \sum_{l=1}^k \sum_{m=1}^k b_{jl}(f) b^{\ast}_{jm}(f) \tau_{lm}, \nonumber \\
    &=& \sum_{m=1}^k |b_{jm}(f)|^2  \tau_{mm}
     +  \sum_{l=1}^k \sum_{m\neq l}^k b_{jl}(f) b^{\ast}_{jm}(f) \tau_{lm}, \nonumber \\
    &=& \sum_{m=1}^k |b_{jm}(f)|^2  \tau_{mm}
     +  \sum_{l=2}^k \sum_{m=1}^{l-1} \bigl(b_{jl}(f) b^{\ast}_{jm}(f)
                               + b_{jm}(f) b^{\ast}_{jl}(f) \bigr) \tau_{lm}, \nonumber \\
    &=& \sum_{m=1}^k |b_{jm}(f)|^2  \tau_{mm}
     + 2\sum_{l=2}^k \sum_{m=1}^{l-1} \bigl(\alpha_{jl}(f) \alpha_{jm}(f)
                                          + \beta_{jl}(f) \beta_{jm}(f) \bigr) \tau_{lm}, 
\end{eqnarray}
where $b^{\ast}_{jk}(f)$ is the complex conjugate of the complex number $b_{jk}(f)$ and $ \alpha_{jm}(f)$ and $\beta_{jm}(f)$ are the real and imaginary part of $b_{jm}(f)$, i.e., $b_{jm}(f) = \alpha_{jm}(f) +i\beta_{jm}(f)$.

For this case, extending Akaike's relative power contribution, 
we define the \textbf{extended absolute power contribution} of $j$-th time series to the fluctuation of $i$-th time series at frequency $f$ by
\begin{eqnarray}
  s_{ij}(f) &=& b_{il}(f) b^{\ast}_{im}(f) \tau_{lm} \nonumber \\
            &=& \left\{  \begin{array}{ll}
                   |b_{ij}(f)|^2  \tau_{jj},  
                   &  j=1,\ldots ,k \\[2mm]
                   \bigl(\alpha_{il}(f) \alpha_{im}(f)
                 + \beta_{il}(f) \beta_{im}(f) \bigr) \tau_{lm}. & 
                   j=k+1, \ldots , k(k+1)/2 ,
                \end{array}\right.
\end{eqnarray}
Here, there is a relationship between $j$ and $(l, m)$, $j=j_l+m$,
where $j_l$ is obtained by the following sequential equation:
\begin{eqnarray}
    j_2  &=& k, \nonumber \\
    j_l  &=& j_{l-1} + l-2, \quad l=3,\ldots ,k.
\end{eqnarray}

Further, the \textbf{extended relative power contribution} of $j$-th time series to the fluctuation of $i$-th time series at frequency $f$ 
is defined by
\begin{eqnarray}
  r_{ij}(f) &=& \frac{b_{il}(f) b^{\ast}_{im}(f) \tau_{lm}}{p_{ii}(f)} \nonumber \\
            &=& \left\{  \begin{array}{ll}
                   \displaystyle\frac{|b_{ij}(f)|^2  \tau_{jj}}{p_{ii}(f)},  
                   &  j=1,\ldots ,k \\[4mm]
                   \displaystyle\frac{\bigl(\alpha_{il}(f) \alpha_{im}(f)
                 + \beta_{il}(f) \beta_{im}(f) \bigr) \tau_{lm}}{p_{ii}(f)}, & 
                   j=k+1, \ldots , k(k+1)/2.
                \end{array}\right.
\end{eqnarray}
Note that for $j=1,\ldots ,k$, $r_{ij}(f)$ is the contribution of the $\tau_{jj}$ which is equivalent to the Akaike's relative power contribution, whereas for $j=k+1,\ldots ,k(k+1)/2$, they are the contribution of the correlated noise corresponding to the covariance 
$\tau_{lm}$. They are real values but are not ncecessarily be positive. 

\newpage
\section{Example of Ship's Data}

\begin{figure}[tbp]
\begin{center}
\includegraphics[width=1.0\textwidth,angle=0,clip=]{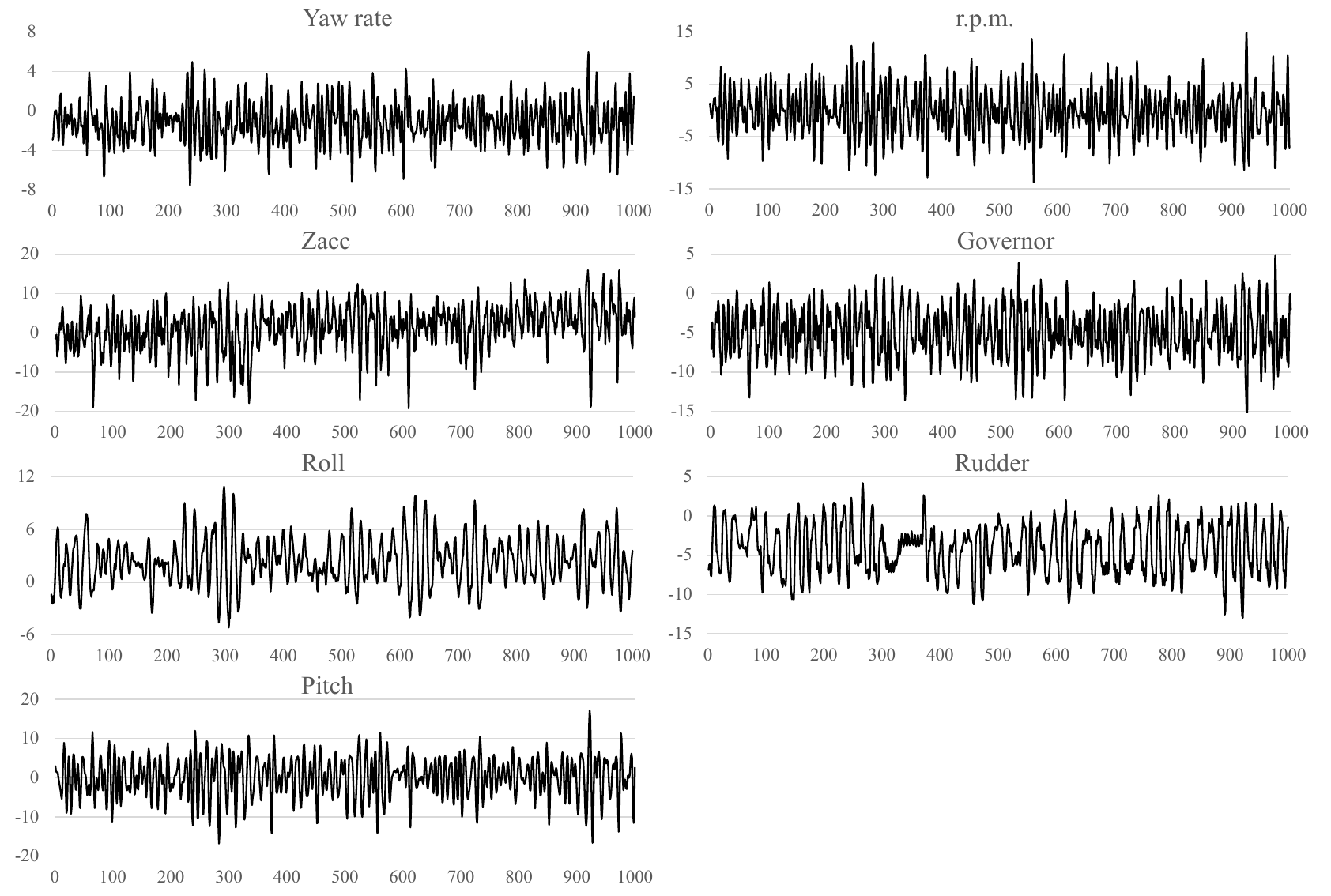}
\end{center}
\caption{Hakusan-Maru ship data. Data length $N=1,000$, sampling inteval $\Delta t=1$sec.}
\label{Fig_Hakusan_data}
\end{figure}

We consider here a 7-variate ship's data, Hakusan-Maru data, consisting of yaw rate, Zacc (vertical acceralation), roll, pitch, engine r.p.m., engine governor and rudder angle shown in Figure \ref{Fig_Hakusan_data} provided by late Prof. Ohtsu of Tokyo University of Marine Science and Technology. Number of observations is $N=1,000$ and sampling interval is 1 second.

\begin{figure}[tbp]
\begin{center}
\includegraphics[width=1.0\textwidth,angle=0,clip=]{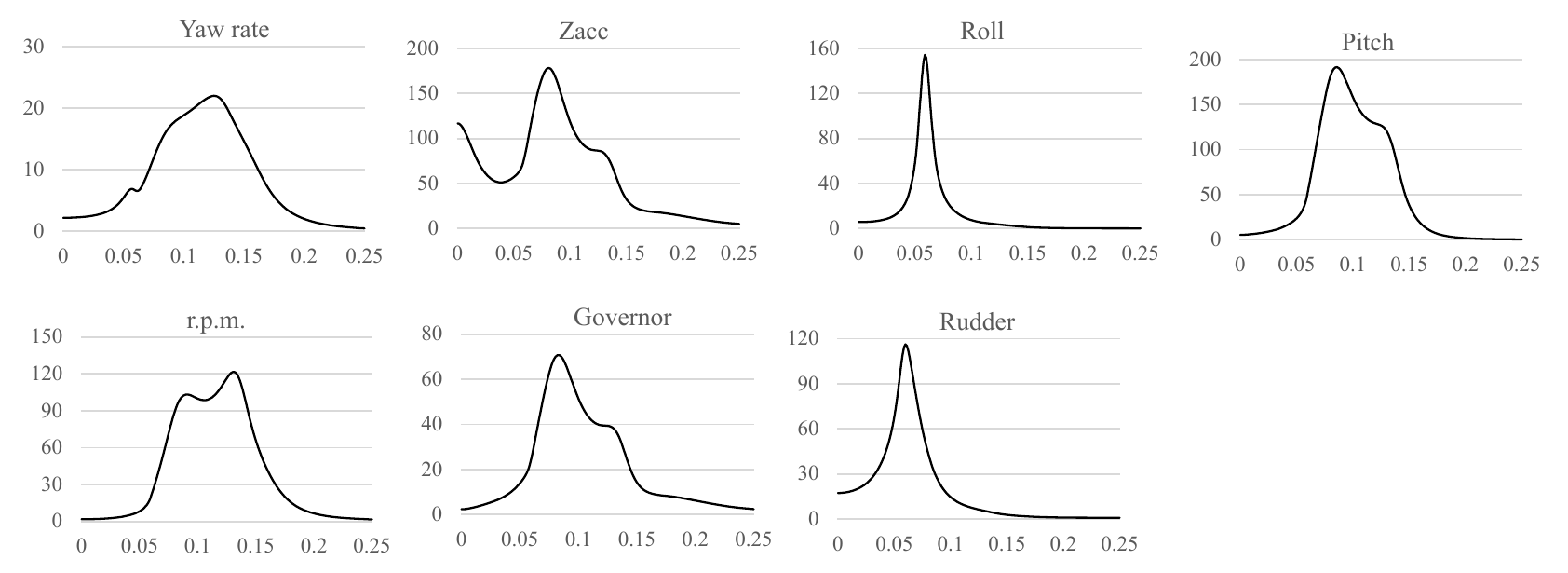}
\end{center}
\caption{Hakusan data.}
\label{Fig_Hakusan_power_spectra}
\end{figure}

Figure \ref{Fig_Hakusan_power_spectra} shows power spectra $p(f)$, $0\leq f \leq 0.25$ (Hz) of 7 variables;
 yaw rate, Zacc, roll, pitch, r.p.m., governor and rudder.
Table \ref{Tab_location of peaks} shows the frequencies of the first and the second
largest peaks (including the inflection point) of each power spectrum.
The frequencies of the maximum peak are classified into three frequencies: about 0.6Hz for roll and rudder, about 0.8Hz for Zacc, pitch, and governor, and 0.125Hz for yaw rate, but the frequencies of the second peaks are concentrated at 0.124, except for yaw rate, which is a very small peak.

\begin{table}[h]
\caption{Frequencies (Hz) of the first and the second peaks of the power spectra.}
\label{Tab_location of peaks}
\vspace{2mm}
\begin{tabular}{lccccccc}
        & Yaw rate & Zacc & Roll & Pitch & r.p.m. & $\!\!$Governor & $\!\!$Rudder \\
\hline
Frequency of heighest peak   & 0.125 & 0.081 & 0.059 & 0.085 & 0.131 & 0.081 & 0.060 \\
Frequency of the second peak & 0.054 & 0.124 & --    & 0.121 & 0.124 & 0.124 & --    \\
\hline
\end{tabular}
\end{table}

\begin{figure}[tbp]
\begin{center}
\includegraphics[width=0.48\textwidth,angle=0,clip=]{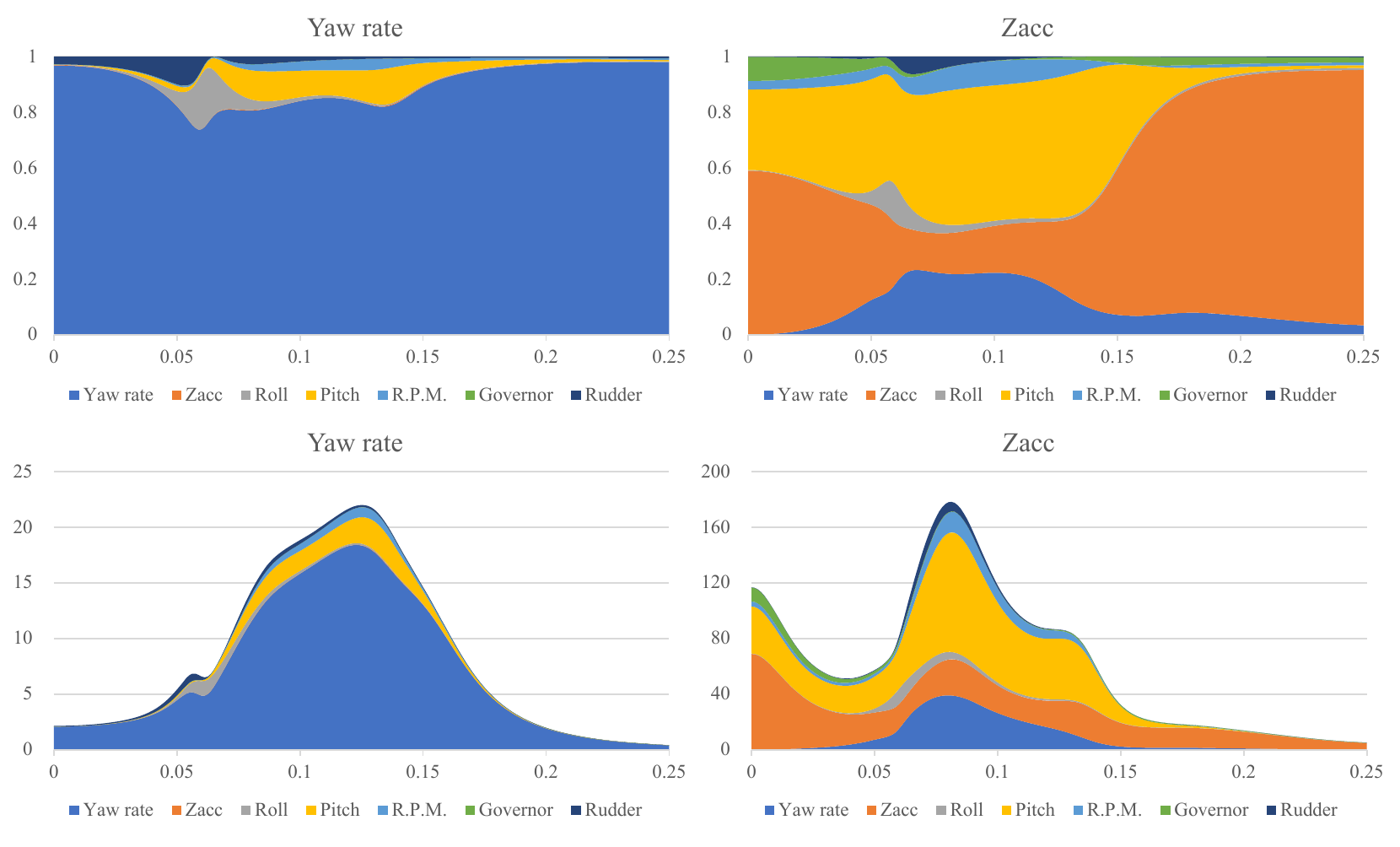}
\includegraphics[width=0.48\textwidth,angle=0,clip=]{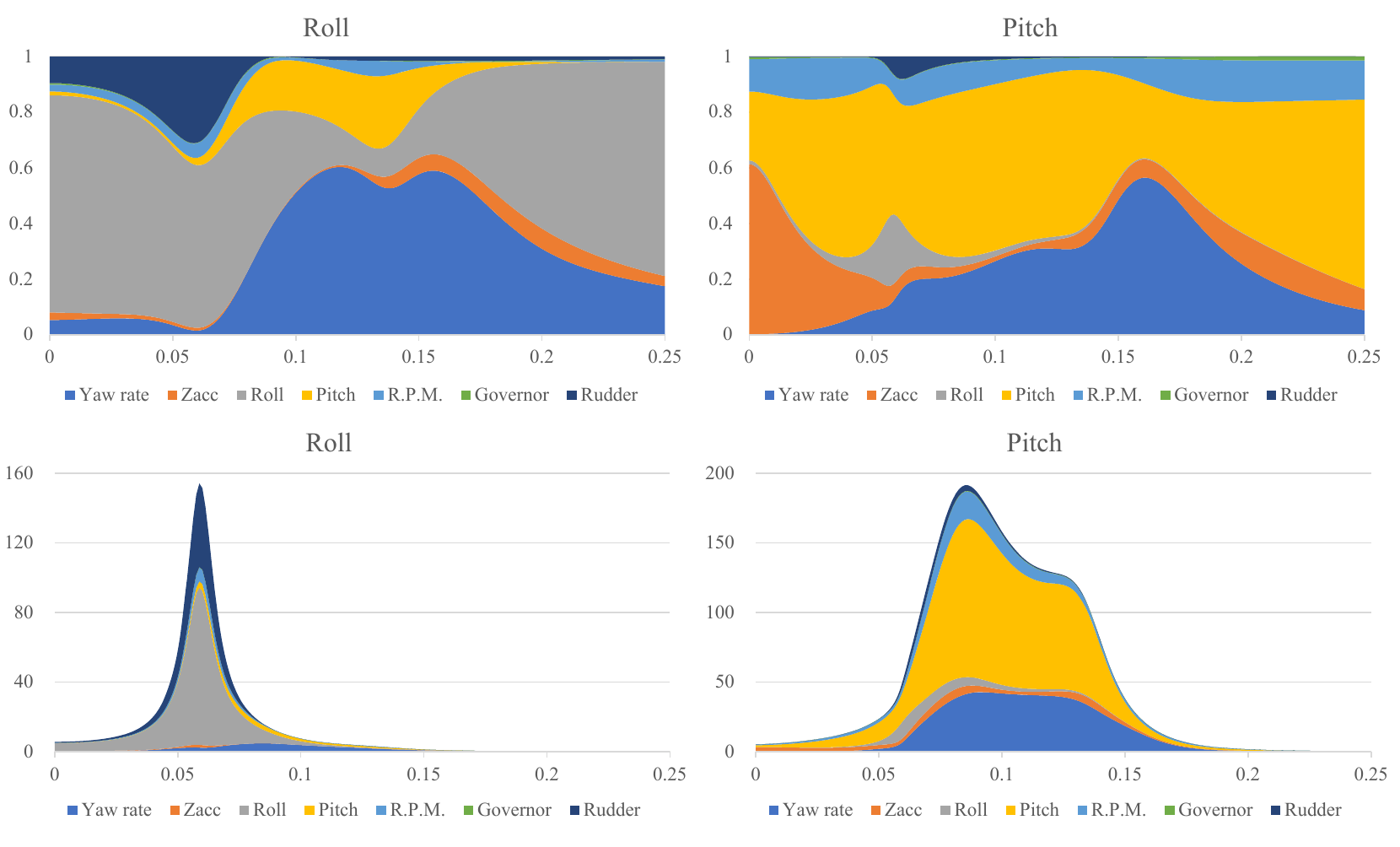}
\includegraphics[width=0.48\textwidth,angle=0,clip=]{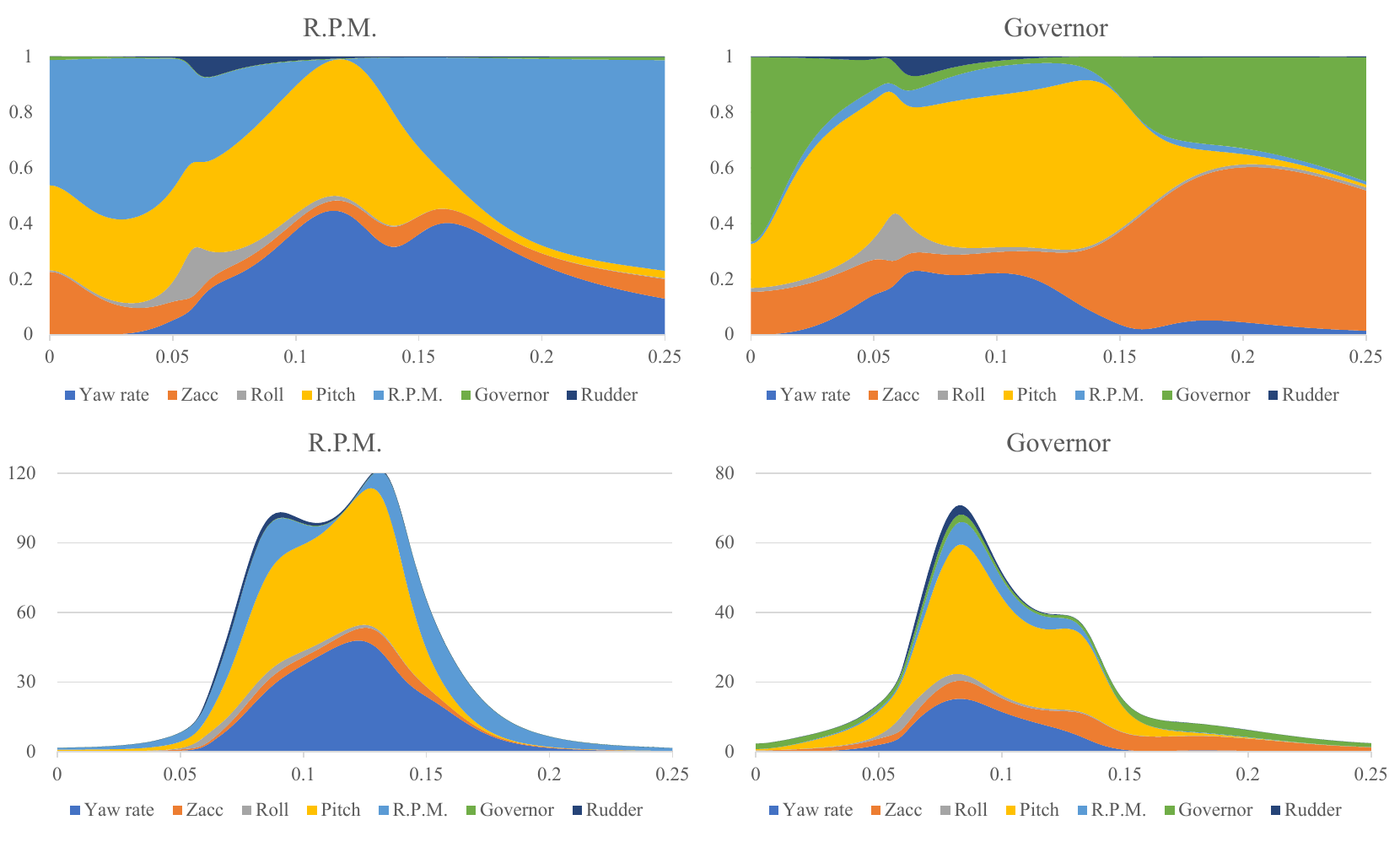}
\includegraphics[width=0.48\textwidth,angle=0,clip=]{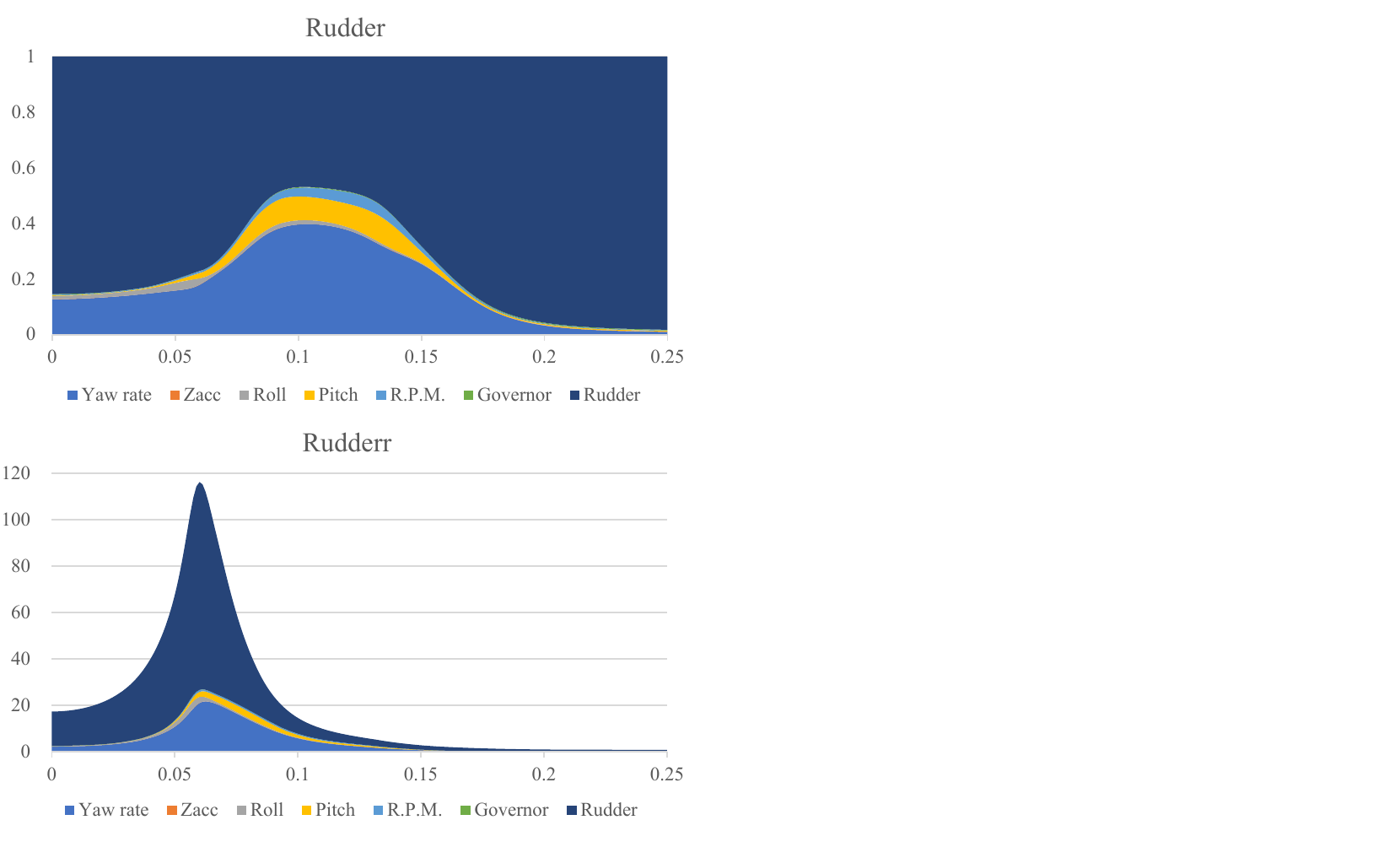}
\end{center}
\caption{Relative power contribution of Hakusan-Maru data.}
\label{Fig_Power-contribution_Hakusan}
\end{figure}

Figure \ref{Fig_Power-contribution_Hakusan} shows Akaike's relative and absolute power
contribution of the Hakusan-Maru data. For each time series, upper panel shows the 
relative power contribution and the lower panel shows the absolute power contribution.

The relative power contribution is a good way to see how much noise contributes to the variation of the time series at each frequency. 
Yaw rate fluctuates over the entire frequency range due to noise directly introduced into the Yaw rate itself.
Rudder is similarly dominated by its own noise, but for frequency $0.05<f<0.15$, it is also strongly influenced by yaw rate.
Yaw rate also has nearly 50\% influence on roll, pitch, and r.p.m. in a certain frequency range.
Pitch shows a strong influence on Zacc, r.p.m., and governor as well as pitch itself.

However, even when the effect of a variable is very strong, if the power spectrum at that frequency is small, the effect on the variability of that variable is not significant. The absolute power contribution allows for an actual effect analysis that takes this into account.
For example, in the case of the roll variation, the influence of roll itself is strongest in the high power region, and the influence of rudder is also found to be about 30\%. Zacc is afffected by pitch, yaw rate and Zacc itself.  Pitch has a strong influence of about 50\% in the major fluctuation regions of Zacc, pitch, r.p.m. and governor, and yaw rate has an influence of about 20--40\% in the major regions of Zacc, pitch, r.p.m. and governor.

\begin{table}[h]
\caption{Correlation matrix of the noise of ship data}\label{Tab_Correlation_hakusan}
\begin{center}
\begin{tabular}{l|rrrrrrr}
       & Yaw rate & Zacc & Roll & Pitch & r.p.m. & Governor & Rudder \\
\hline
Yaw rate &  1.000  &    0.000  &    0.000  &   -0.197  &    0.029  &   -0.021  &    0.001\\
Zacc     &  0.000  &    1.000  &    0.064  &    0.075  &   -0.057  &    0.084  &   -0.008\\
Roll     &  0.000  &    0.064  &    1.000  &   -0.068  &   -0.070  &   -0.002  &   -0.041\\
Pitch    & -0.197  &    0.075  &   -0.068  &    1.000  &   -0.144  &    0.111  &   -0.002\\
r.p.m.   &  0.029  &   -0.057  &   -0.070  &   -0.144  &    1.000  &   -0.020  &   -0.026\\
Governor & -0.021  &    0.084  &   -0.002  &    0.111  &   -0.020  &    1.000  &   -0.082\\
Rudder   &  0.001  &   -0.008  &   -0.041  &   -0.002  &   -0.026  &   -0.082  &    1.000
  \end{tabular} 
\end{center}
\end{table}
The correlation matrix of the noise $V$ is shown in Table \ref{Tab_Correlation_hakusan}.
From this correlation matrix, we can see that the absolute value of the correlation coefficient of the noise is small, at most 0.197. Many correlation coefficients are statistically significant since the variance of the sample correlation coefficients for time series can be approximated by $1/\sqrt{N+2}$=0.0316 (Box and Jenkins (2015)). However, in applying Akaike's relative power contribution, we will be able to consider that the correlations are sufficiently small except for the correlatted noises (pitch and yaw rate), (r.p.m. and pitch), and (governor and pitch).

Therefore, the assumption that the variance-covariance matrix of the noise is diagonal is relatively reasonable for the Hakusan-Maru data, and the above analysis using Akaike's power contribution ratio is appropriate. 

\begin{figure}[tbp]
\begin{center}
\includegraphics[width=0.48\textwidth,angle=0,clip=]{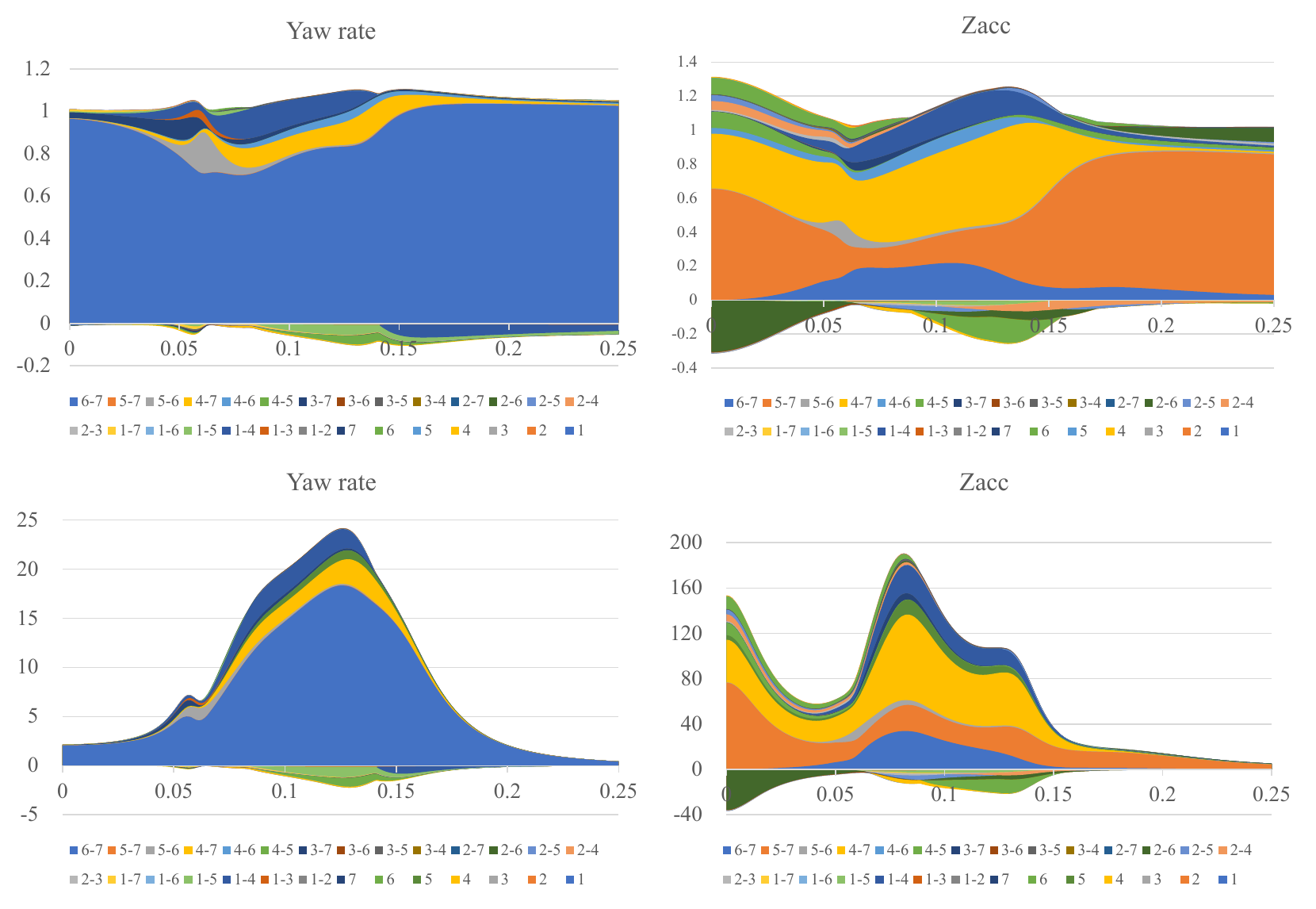}
\includegraphics[width=0.48\textwidth,angle=0,clip=]{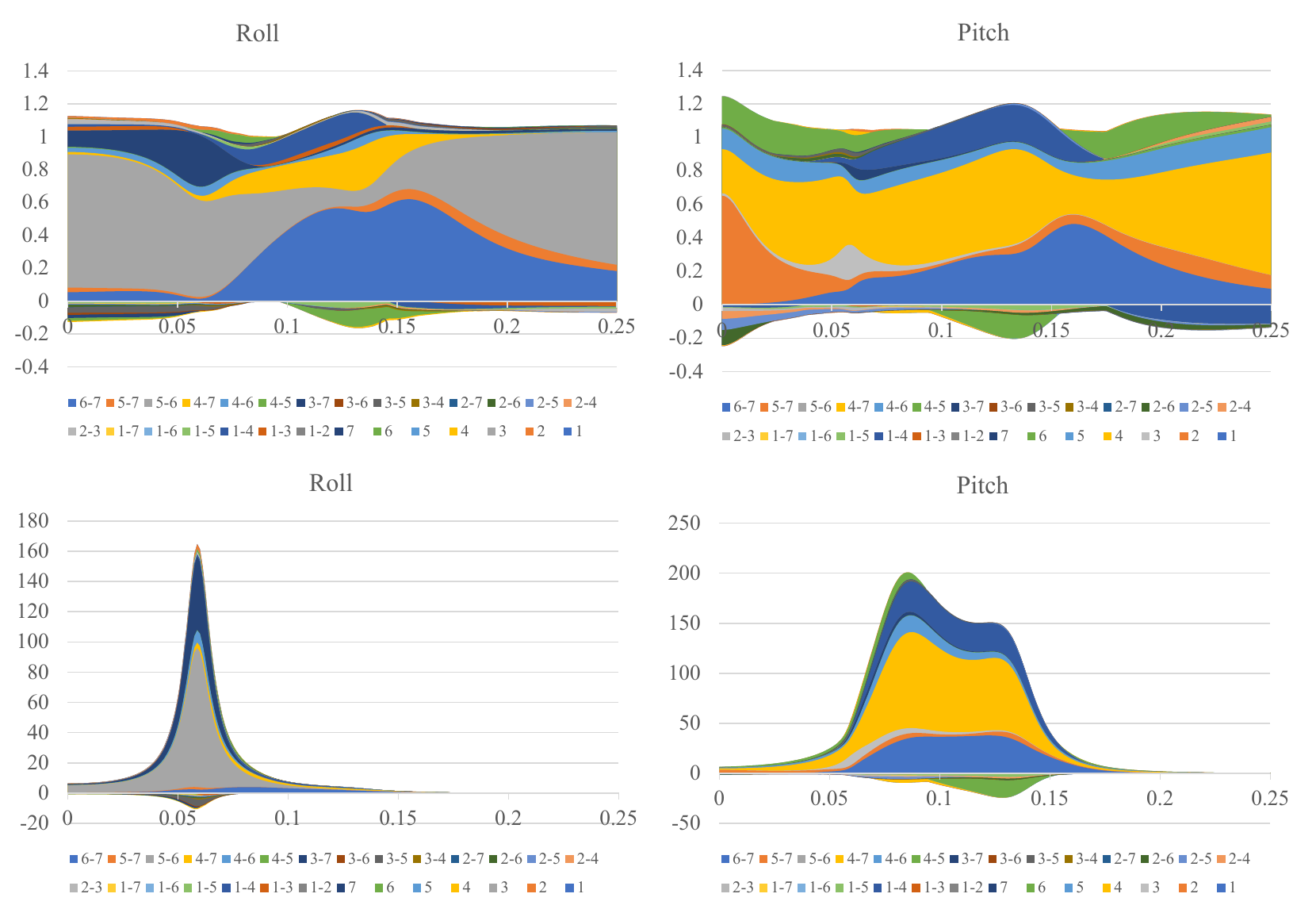}
\includegraphics[width=0.48\textwidth,angle=0,clip=]{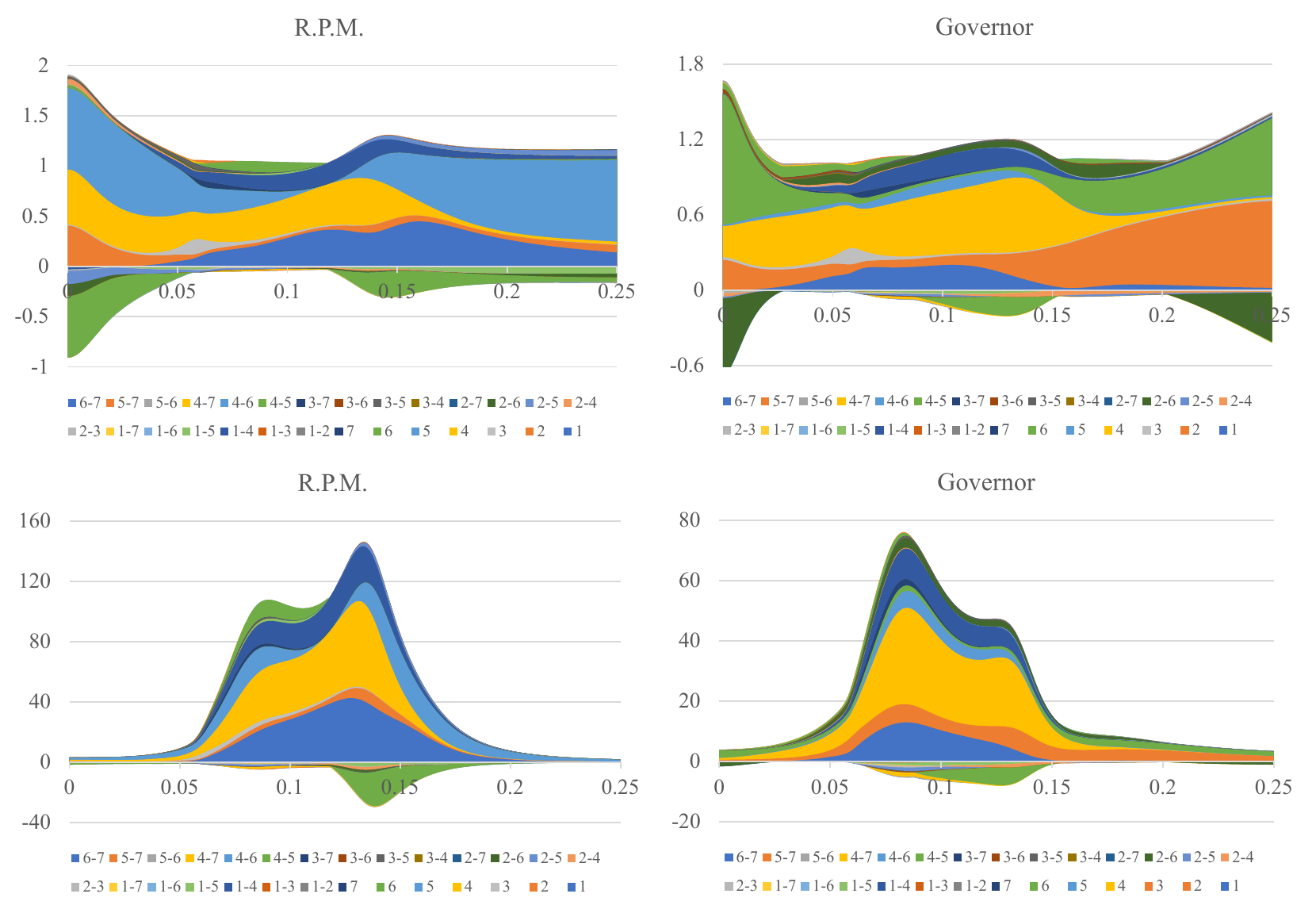}
\includegraphics[width=0.48\textwidth,angle=0,clip=]{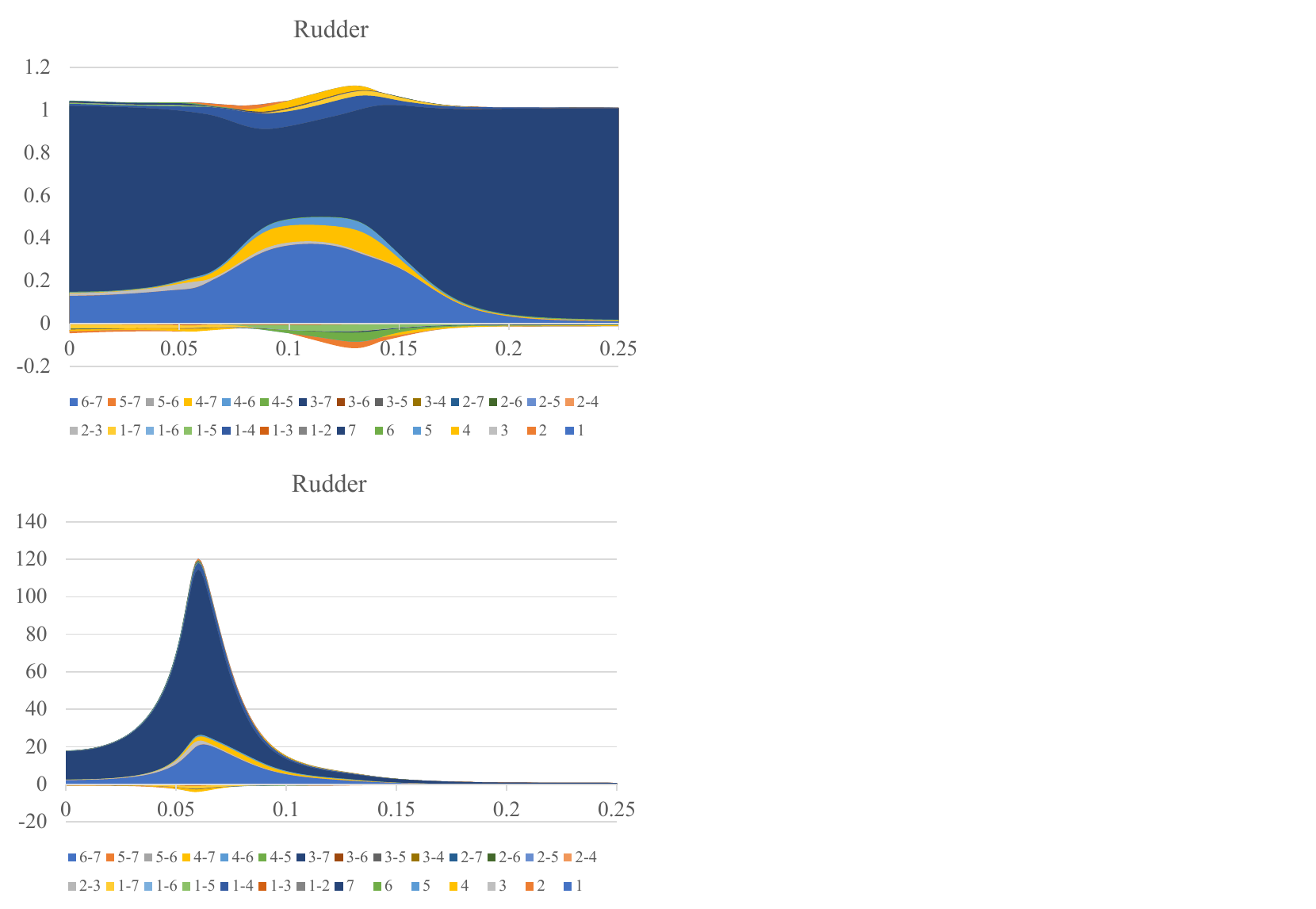}
\end{center}
\caption{Extended relative power contribution of Hakusan-Maru data.}
\label{Fig_Extended power-contribution_Hakusan}
\end{figure}

Figure \ref{Fig_Extended power-contribution_Hakusan} shows the results of calculating the extended power contribution for reference. In some parts, the independent noise is replaced by correlated noise, but in general, almost the same decomposition is obtained, confirming that good results are obtained in the analysis assuming uncorrelatedness of the noise in the case of this data.

\newpage
\section{Example of GDP Data}

\begin{figure}[h]
\begin{center}
\includegraphics[width=0.8\textwidth,angle=0,clip=]{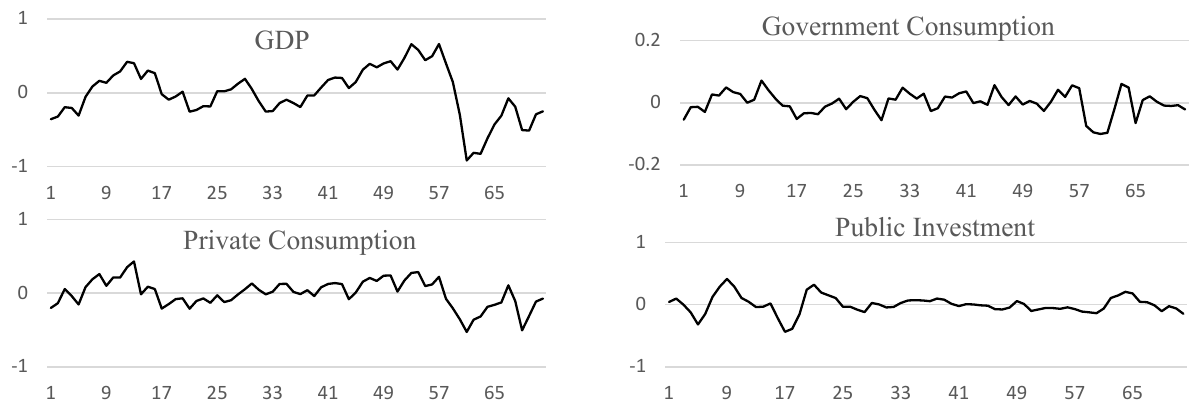}
\end{center}
\caption{GDP data.}
\label{Fig_GDP4_data}
\end{figure}

In this section, we consider quarterly four-variate macroeconomic data for Japan (GDP, private consumption, government consumption, public investment, 19XX-202XX, N=72, AR components of seasonally adjusted data by DECOMP) shown in Figure \ref{Fig_GDP4_data}.

\begin{figure}[tbp]
\begin{center}
\includegraphics[width=0.48\textwidth,angle=0,clip=]{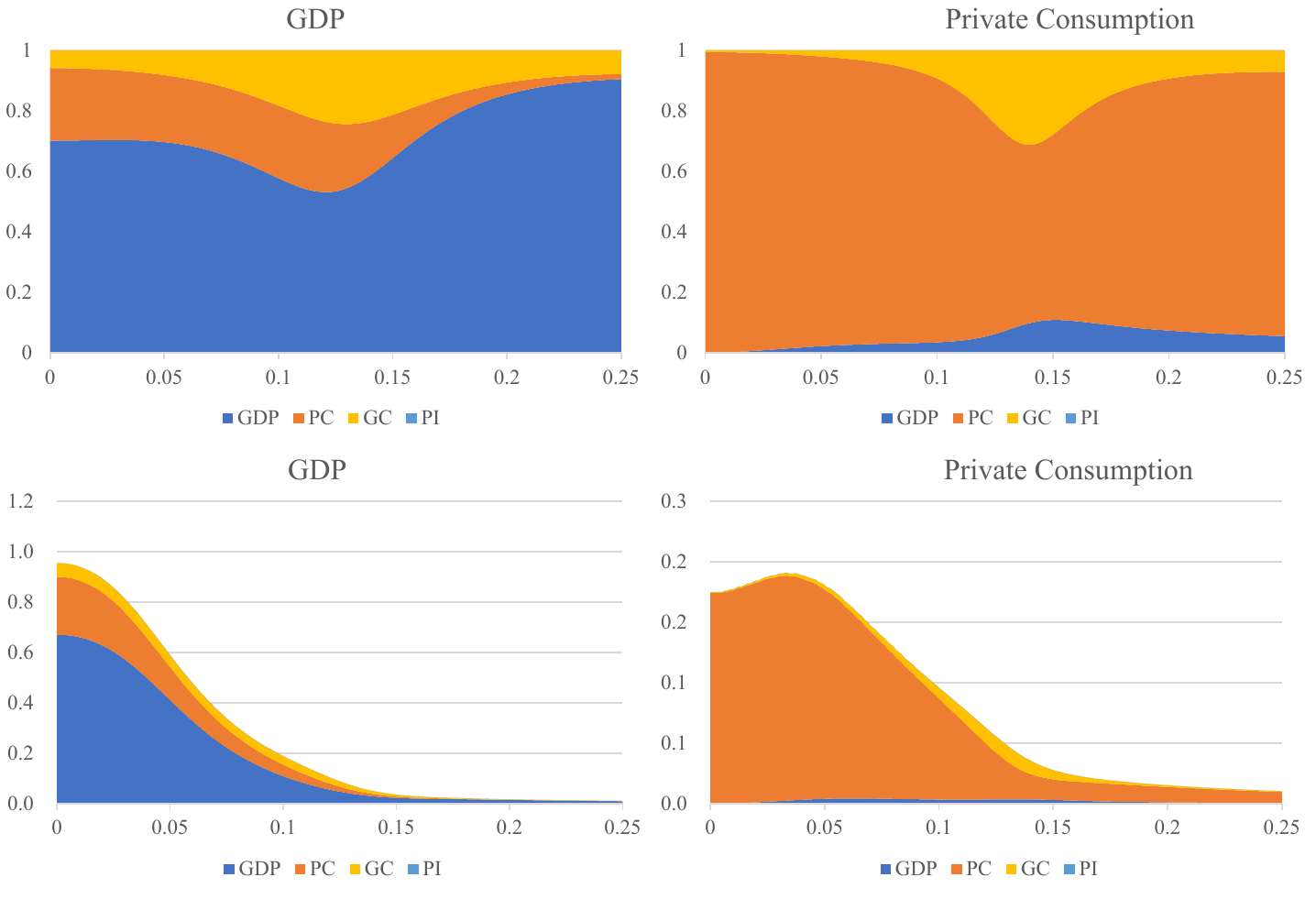}
\includegraphics[width=0.48\textwidth,angle=0,clip=]{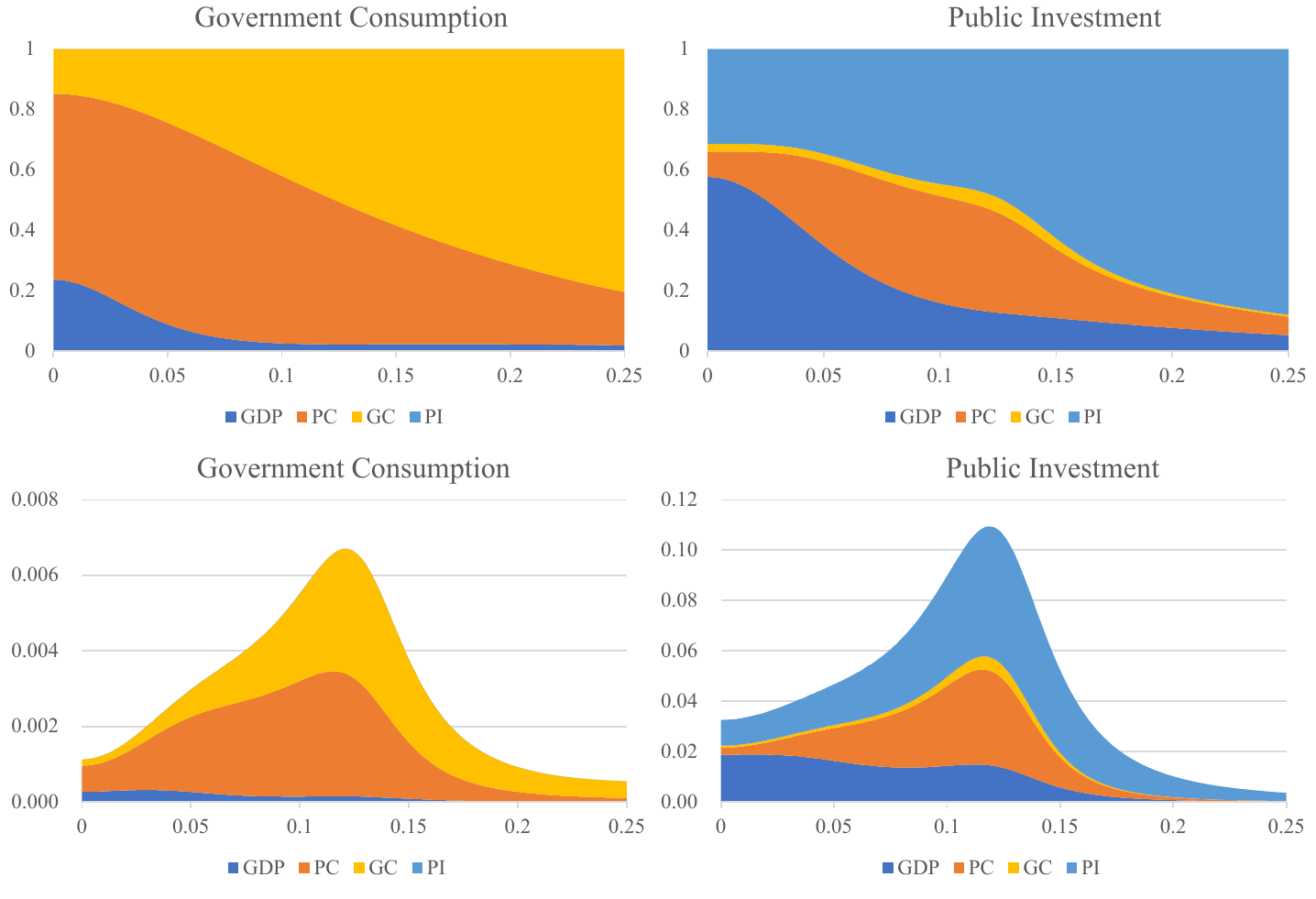}
\end{center}
\caption{Relative power contribution of GDP data.}
\label{Fig_Power-contribution_GDP4}
\end{figure}

Figure \ref{Fig_Power-contribution_GDP4} shows Akaike's power contribution ($0<f<0.25$Hz) obtained assuming that the variance-covariance matrix of the noise is diagonal. 
For GDP, there is about 20\% contribution from PC in the range $0<f<0.1$. The variation of private consumption is almost entirely due to its own contribution in the low frequency range, $f<0.1$.
For government consumption, the contribution from private consumption is about 50\% at around 0.125 of the spectral peak, but private consumption is dominant at lower frequencies and government consumption is dominant at higher frequencies.
For public investment, the GDP contribution is about 20\% and the private consumption contribution is about 30\%.

However, Table \ref{Tab_correlation matrix of GDP data} shows the correlation coefficient matrix of the noise, and the absolute values of all correlation coefficients exceed the sample variance (0.116), especially $r_{21}$, $r_{31}$, and $r_{32}$ are very large. This suggests that the usual power contribution may not be sufficient to capture the variability characteristics of this data.
\begin{table}[h]
\caption{Correlation matrix of the noise}\label{Tab_correlation matrix of GDP data}
\begin{center}
  \begin{tabular}{l|cccc}
     &   GDP    &   PC     &    GC    &   PI   \\
\hline
 GDP & 1.00000  & 0.84575  & 0.43107  & 0.22950\\
 PC  & 0.84575  & 1.00000  & 0.50969  & 0.27136\\
 GC  & 0.43107  & 0.50969  & 1.00000  & 0.13831\\
 PI  & 0.22950  & 0.27136  & 0.13831  & 1.00000
  \end{tabular}
 \end{center}
\end{table}

\begin{figure}[tbp]
\begin{center}
\includegraphics[width=0.48\textwidth,angle=0,clip=]{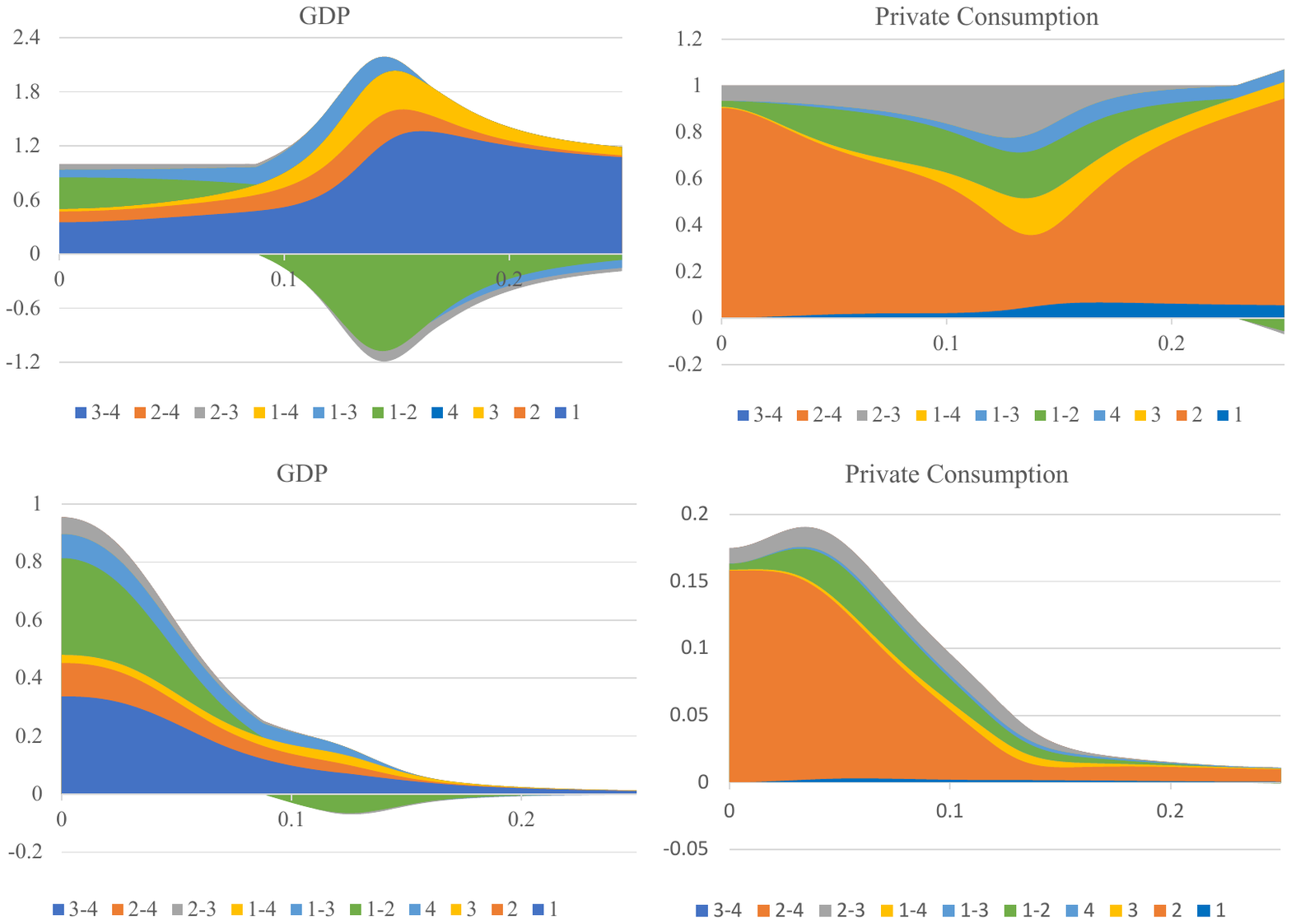}
\includegraphics[width=0.48\textwidth,angle=0,clip=]{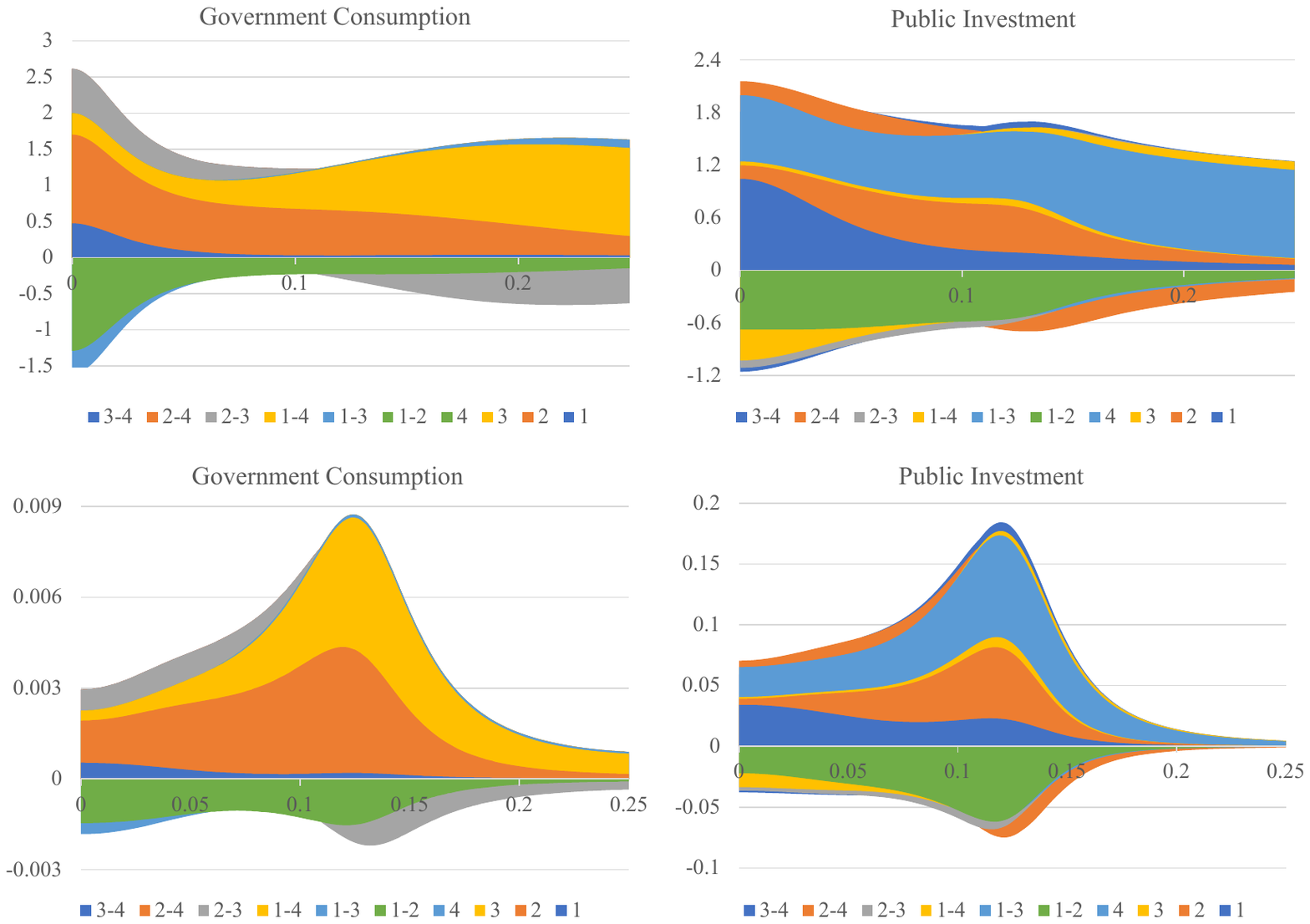}
\end{center}
\caption{Extended relative power contribution of GDP data. Labels 1, 2, 3, 4 denote
GDP, Private consumption, Government consumption and Public investment, respectively.}
\label{Fig_Extended-power-contribution_GDP4}
\end{figure}

Figure \ref{Fig_Extended-power-contribution_GDP4} shows the extended power contribution of the GDP data. In this case, unlike the usual relative power contribution in Figure \ref{Fig_Power-contribution_GDP4}, the appearance of negative values and values above 1 is the main feature. In the case of GDP, GC, and PI, the contributions of the correlated noise (GDP+PC) are negative, with an equal amount of values above 1 appearing on the positive side. In the case of PI, the effects of (GDP+PI) and (PC+PI) are also negative, which means that the power spectrum is reduced by the negative contribution of the correlated noise.

\begin{figure}[tbp]
\begin{center}
\includegraphics[width=0.60\textwidth,angle=0,clip=]{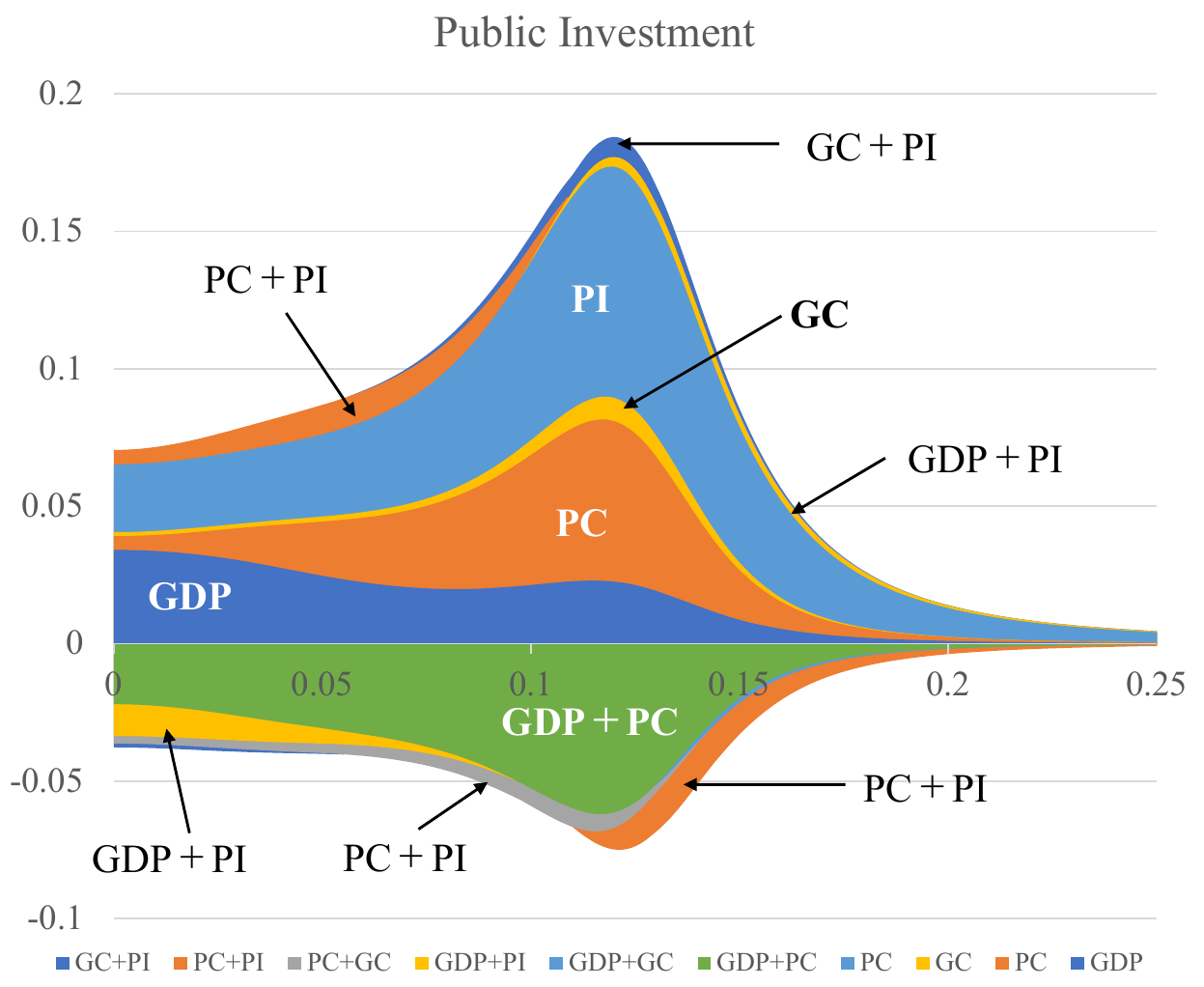}

\end{center}
\caption{Extended relative power contribution to PI.}
\label{Fig_Extended-power-contribution_GDP4-3}
\end{figure}

Figure \ref{Fig_Extended-power-contribution_GDP4-3} is an enlarged plot of the absolute power contribution of the PI in the lower right corner of Figure \ref{Fig_Extended-power-contribution_GDP4-3}.
The contribution from correlated noise (PC+PI) is positive for $f<0.1$ but negative for $f>0.1$. On the other hand, the contribution from (GDP+PI) is the opposite of this, being negative for $f<0.1$ and positive for $f>0.1$.

\newpage
\section{Simulation analysis}

In this section, we revisit the findings obtained by the power contribution analysis from a simulation perspective using the GDP data.
Here we first obtain the realization of noise input $e_n=(e_n^{\mbox{\tiny GDP}},e_n^{\mbox{\tiny PC}},e_n^{\mbox{\tiny GC}},e_n^{\mbox{\tiny PI}})^T$ by
\begin{eqnarray}
  e_n = y_n - \sum_{j=1}^m A_j y_{n-j}.
\end{eqnarray}
Then the effect of the noise input to a certain series is obtained by
\begin{eqnarray}
  y_n^{\delta} = \sum_{j=1}^m A_j y_n^{\delta} + e_n^{\delta},\quad  \mbox{for }n=37,\ldots ,72,
\end{eqnarray}
where $\delta$ ($=1,\ldots ,4$) is either of GDP, PC, GC or PI and as the initial values we set $y_n^{\delta} = y_n$ for $n=1,\ldots ,m$.

\begin{figure}[tbp]
\begin{center}
\includegraphics[width=0.48\textwidth,angle=0,clip=]{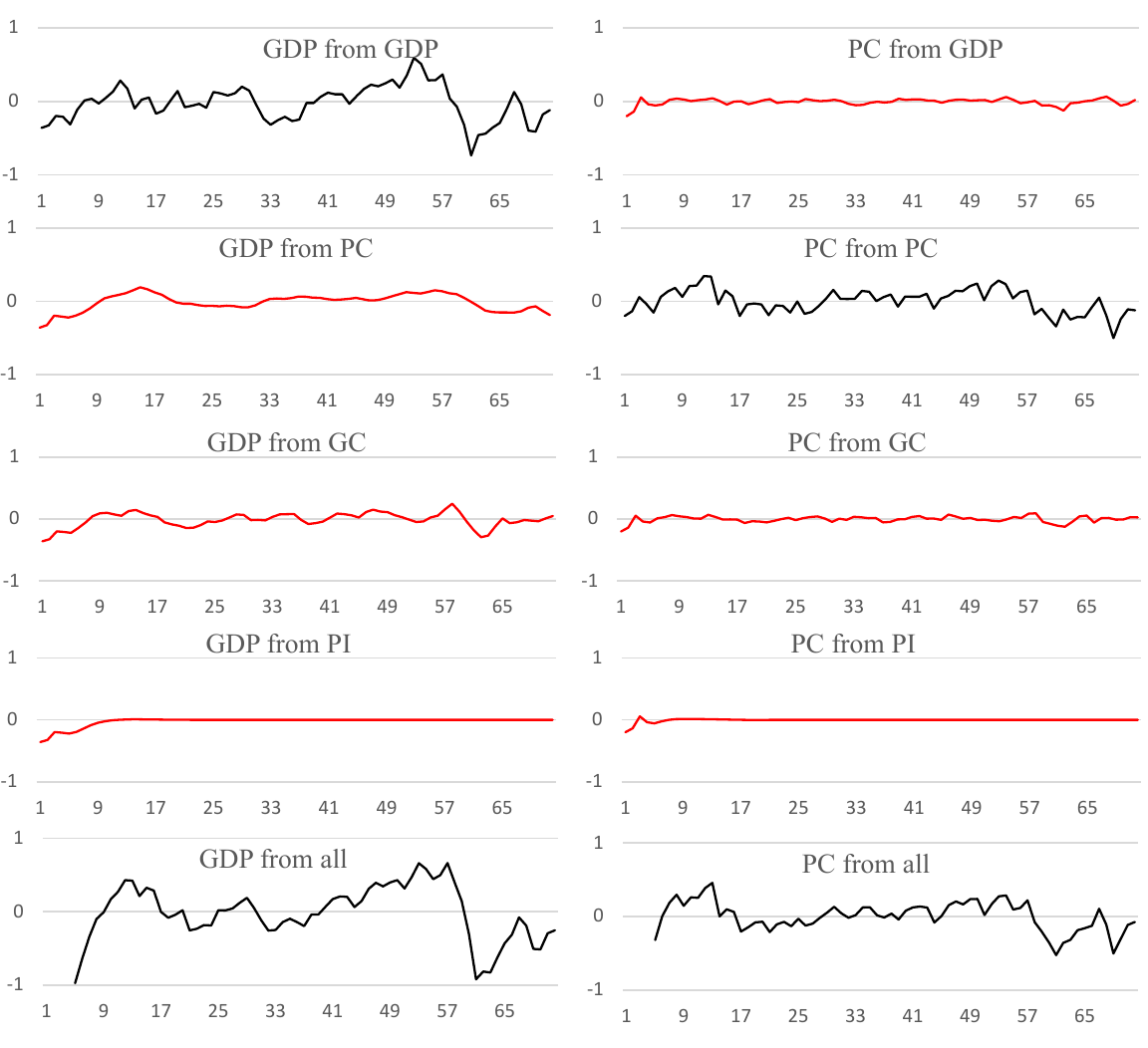}
\includegraphics[width=0.48\textwidth,angle=0,clip=]{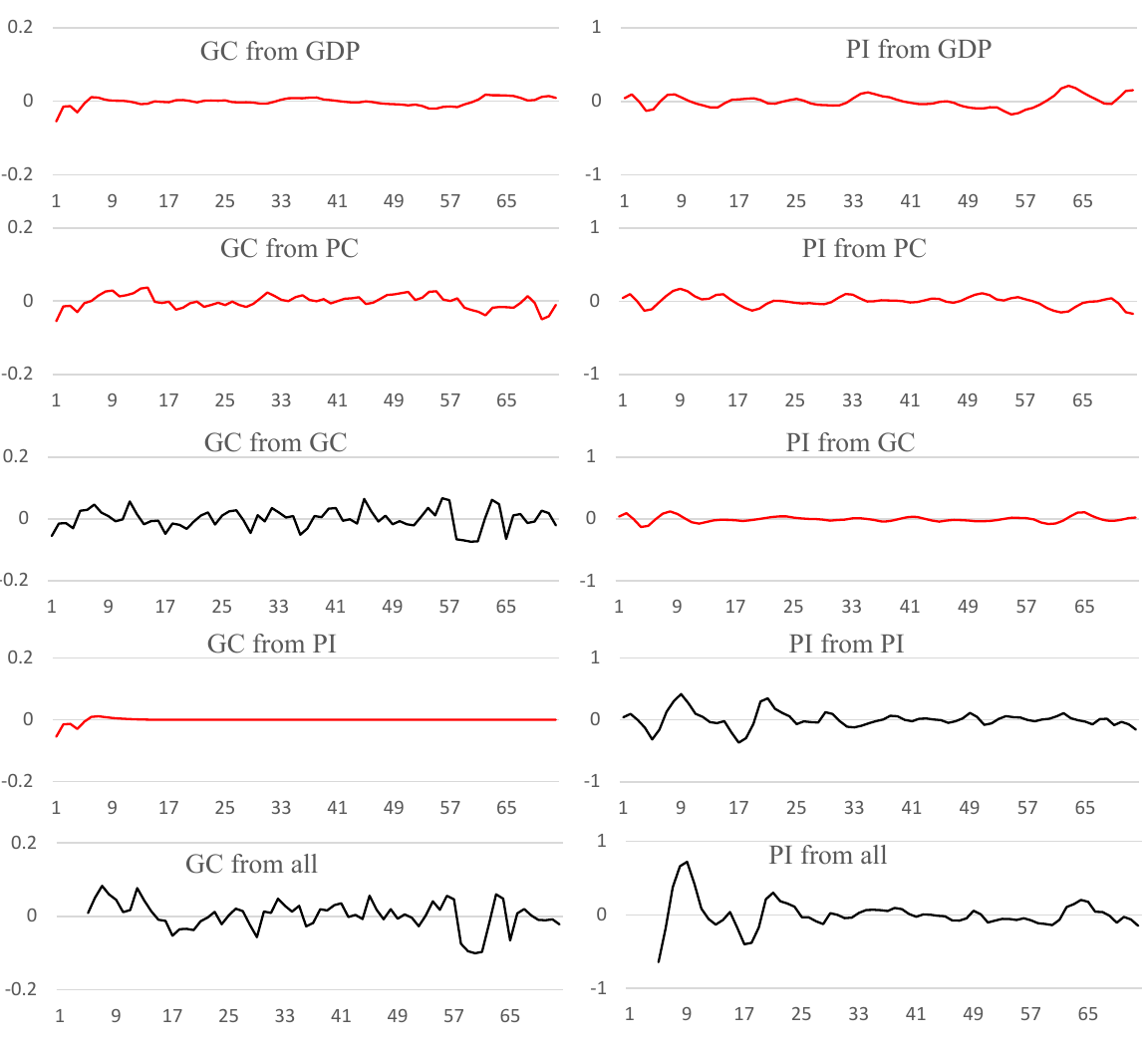}
\end{center}
\caption{Simulation analysis of GDP data.}
\label{Fig_Simulation_GDP4_1}
\end{figure}

In Figure \ref{Fig_Simulation_GDP4_1}, the results for the GDP, PC, GC, and PI components are shown from left to right. For each component, the contributions of GDP, PC, GC, and PI and the sum of these contributions are shown from top to bottom.
In the case of GDP, it can be seen that most of the variation is caused by the noise of GDP, while the noise of PC causes the long-period component and the noise of GC causes part of the short-period component, and the effect of PI is almost negligible. This is consistent with the power contribution analysis shown in Figure \ref{Fig_Power-contribution_GDP4}.
In the case of PC, most of the variation is caused by PC noise, but the effect of GC noise is also observed.
In the case of GC, the contribution of GS noise accounts for more than half of the variation, while the contribution of GDP is observed on the long-period side, and that of PC on the short-period variation.
In the case of PI, GDP, PC, GC, and PI all contribute to the fractuation of PI.
These results are also consistent with the power contribution analysis results in Figure \ref{Fig_Power-contribution_GDP4}.

\begin{figure}[h]
\begin{center}
\includegraphics[width=0.48\textwidth,angle=0,clip=]{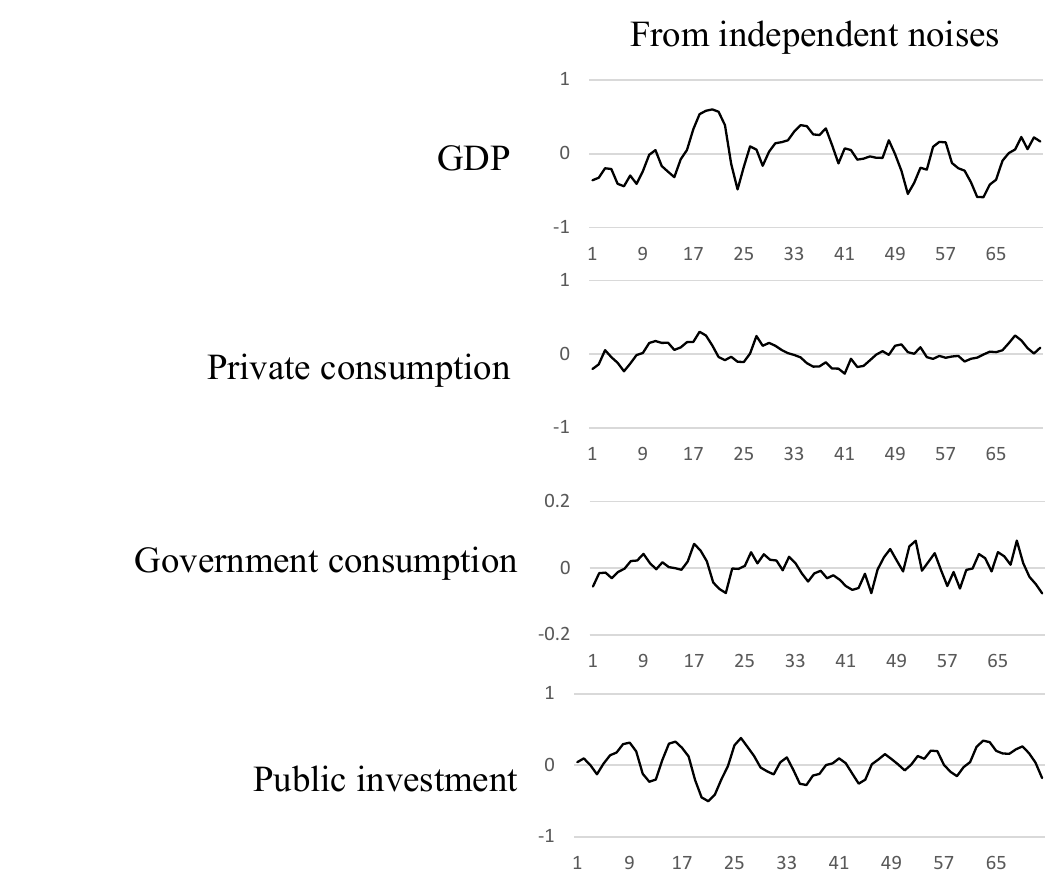}
\includegraphics[width=0.48\textwidth,angle=0,clip=]{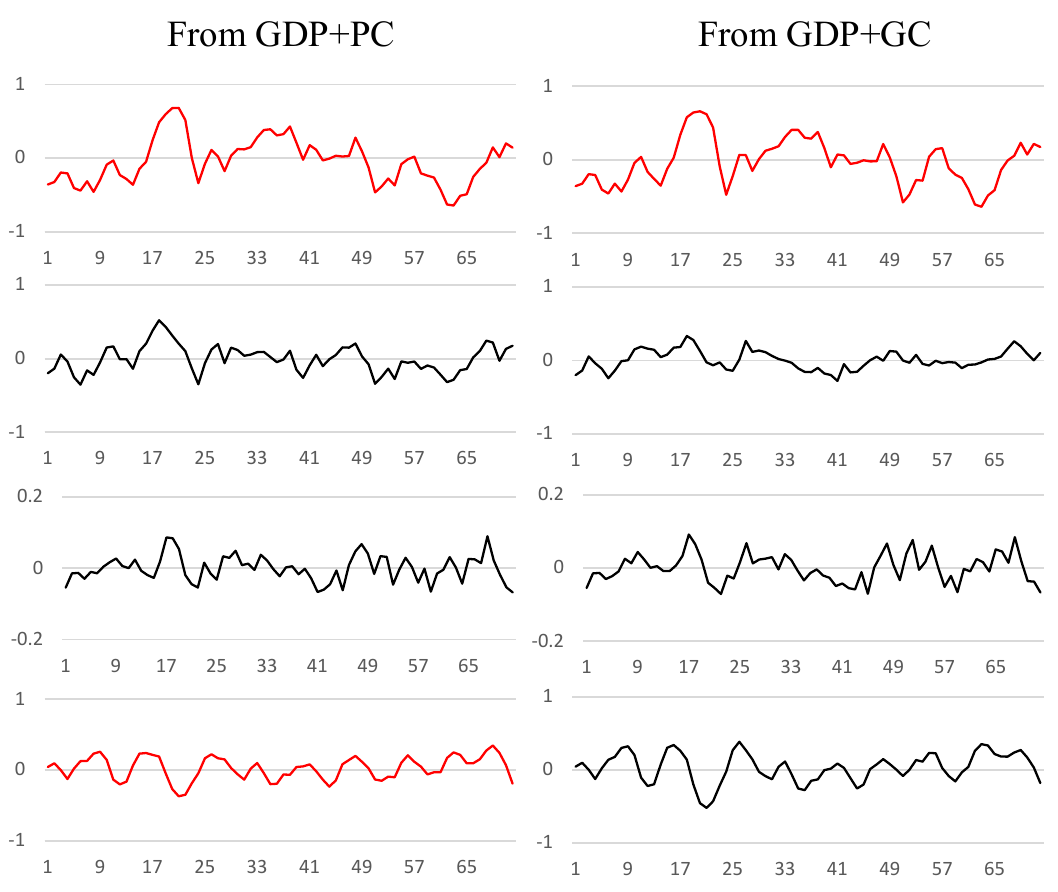}

\vspace{5mm}
\includegraphics[width=0.48\textwidth,angle=0,clip=]{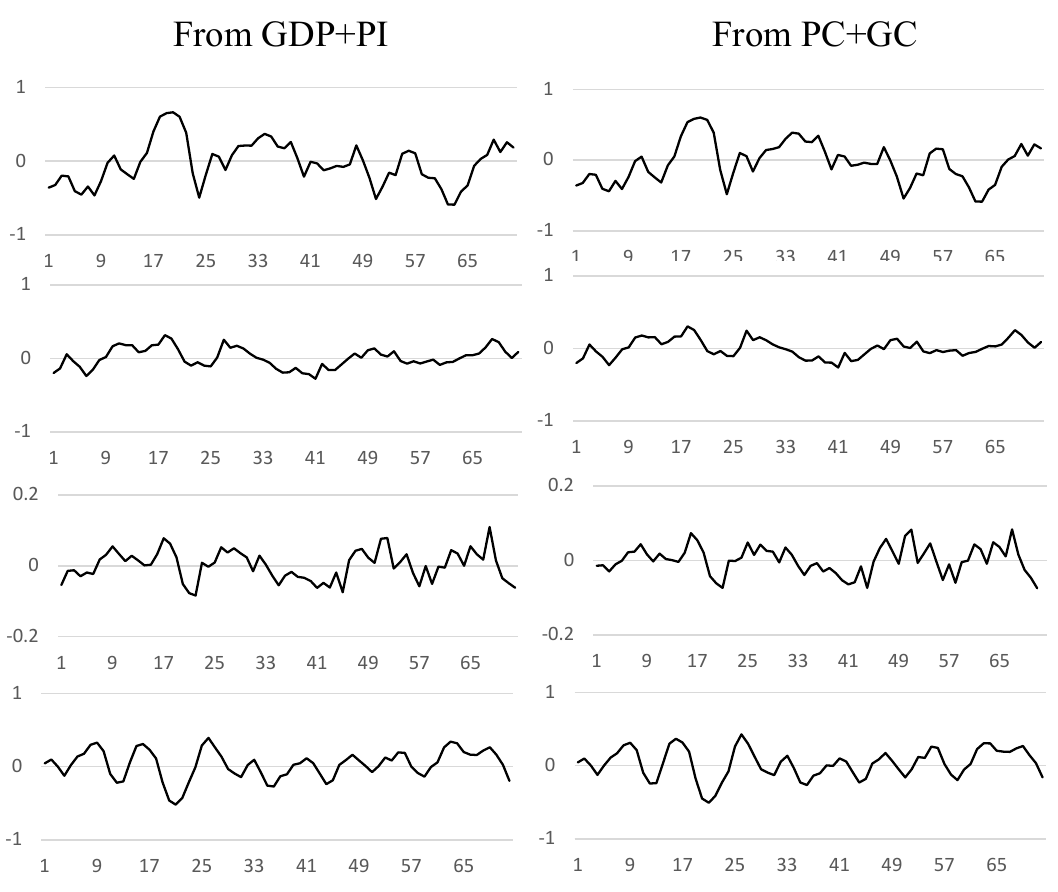}
\includegraphics[width=0.48\textwidth,angle=0,clip=]{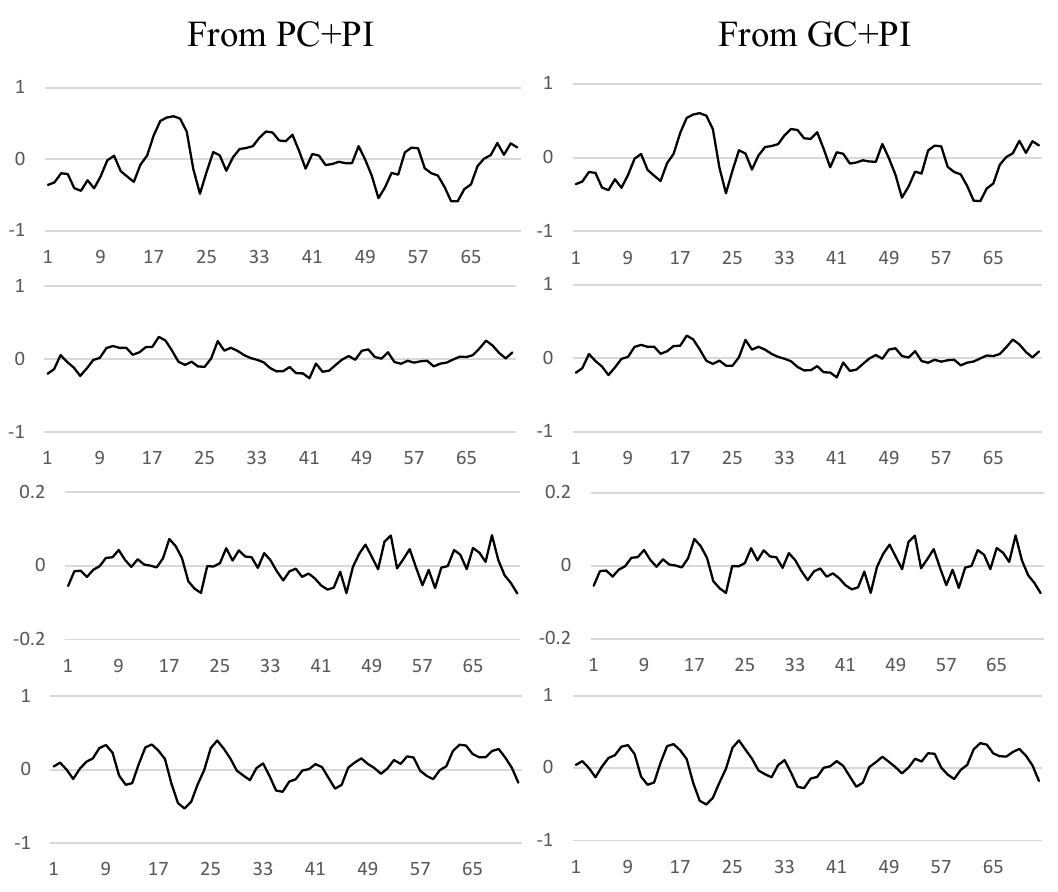}
\end{center}
\caption{Simulation analysis of GDP data.}
\label{Fig_Simulation_GDP4_2}
\end{figure}

Figure \ref{Fig_Simulation_GDP4_2} shows a 4-variate time series (GDP, PC, GC, and PI from top to bottom) generated using pseudo-random numbers.
The upper four plots on the left side are the case where only four independent noises (GDP, PC, GC, and PI) were added. The middle of the upper 4 row shows the simulation results when the correlated noise of GDP and PC is added in addition to these four independent noises, and four right side plots show the simulation results when the correlated noise of GDP and GC is added in addition to the four independent noises.
Similarly, the four columns in the bottom half show, from left to right, the simulation results when correlated noise of (GDP+PI), (PC+GC), (PC+PI), and (GC+PI) are added to the four independent noises, respectively.

Although difficult to see visually, the addition of correlated noise between GDP and PC (GDP+PC) results in smaller fluctuations in PI but larger fluctuations in GDP and PC compared to the case of independent noise alone.
In the case of (GDP+GC), the variation of GC is slightly larger. In other cases, there is no clear difference.

Since it is difficult to find clear differences in a simulation of a short time series of $N=72$, we generated 10,000 time series of length 1000, and the results for the mean and standard deviation are shown in Table \ref{Tab_simulation}. The addition of correlated noise between GDP and PC, labeled (1+2), reduces the variance of PI by 23\%, as indicated by the green letters, while the variance of PC is 3.3 times larger. The variance of GDP also increases by 31\%.
The correlated noise between GDP and GC, labeled (1+3), increases the change in GDP by 18\%, while PC and PI also increase slightly, by 8.5\% and 3.5\%, respectively.
The correlation noise between GDP and PI, labeled (1+4), increases the variance of GDP, PC, and GC by 5.3\%, 15.7\%, and 13.4\%, respectively.
Some influence of PC and GC, labeled (2+3), and PC and PI, labeled (2+4), on PI is also observed, but the increase in variance is only 1.6\%.

\begin{table}[h]
\noindent
\caption{Mean and the variance of the varainces of time series}\label{Tab_Simulation of GDP data}\label{Tab_simulation}
 Mean
\begin{center}
  \begin{tabular}{l|ccccccc}
  & (1,2,3,4)  & (1+2) & (1+3) & (1+4) & (2+3) & (2+4) & (3+4) \\
\hline
GDP & 0.077257 & {\color{red}0.101473} & {\color{red}0.091400} & {\color{red}0.081338} & 0.077257 & 0.077257 & 0.077257\\
PC  & 0.011133 & {\color{red}0.036870} & {\color{red}0.012081} & {\color{red}0.012886} & 0.011133 & 0.011133 & 0.011133\\
GC  & 0.001109 & 0.001134 & 0.001198 & {\color{red}0.001258} & 0.001109 & 0.001109 & 0.001109\\
PI  & 0.025985 & {\color{green}0.019983} & {\color{red}0.026887} & 0.025859 & {\color{red}0.026394} & {\color{red}0.026399} & 0.025985\\
\hline
\end{tabular}
\end{center}
\noindent 
100 times of the standard deviation
\begin{center}
  \begin{tabular}{l|ccccccc}
  & (1,2,3,4)  & (1+2) & (1+3) & (1+4) & (2+3) & (2+4) & (3+4) \\
\hline
GDP & 0.650517 & {\color{red}0.985340} & {\color{red}0.790528} & {\color{red}0.693810} & 0.650517 & 0.650517 & 0.650517\\
PC  & 0.079129 & {\color{red}0.283344} & {\color{red}0.084028} & {\color{red}0.094602} & 0.079129 & 0.079129 & 0.079129\\
GC  & 0.005959 & {\color{red}0.006204} & {\color{red}0.006429} & {\color{red}0.006930} & 0.005959 & 0.005959 & 0.005959\\
PI  & 0.190314 & {\color{green}0.146561} & {\color{red}0.196088} & {\color{red}0.189435} & {\color{red}0.193996} & {\color{red}0.193257} & 0.190314\\
\hline
\end{tabular}
\end{center}
\end{table}

\section{Concluding Remarks}
Although Akaike's relative power contribution is a very useful tool for analysis of multivariate dynamic systems with feedback, this method is not applicable to the time series with significant correlations of the noise inputs. In order to address this problem, we defined a extended power contribution that can be applied to general systems. 

Numerical examples with ship data show that this extended relative power contribution is similar to the Akaike's relative power contribution when the correlation between the noises is low.
On the other hand, the numerical example with the GDP data shows that some of the correlated noise has a negative effect on the power spectrum, which has the effect of reducing the power spectrum of the variability of the time series.
Simulation experiments also confirmed that the variance  may indeed be reduced by negative contribution of correlated noises.

\end{document}